\documentclass[12pt]{article}

\usepackage[a4paper,top=2.5cm,bottom=2.5cm,left=2.5cm,right=2.5cm]{geometry}
\usepackage[T1]{fontenc}
\usepackage[utf8]{inputenc}
\usepackage{amsmath,amssymb,amsthm}
\usepackage{newtxtext,newtxmath}
\usepackage{bm}
\usepackage{authblk}
\usepackage[round,authoryear]{natbib}
\usepackage{graphicx,booktabs,array,tabularx}
\usepackage{enumitem,setspace,microtype,needspace}
\usepackage{xcolor}
\usepackage[hidelinks]{hyperref}
\usepackage{lineno}
\usepackage{multirow}

\newtheorem{lemma}{Lemma}
\newtheorem{theorem}{Theorem}
\newtheorem{corollary}{Corollary}

\theoremstyle{remark}

\DeclareMathOperator*{\argmax}{arg\,max}

\DeclareMathOperator{\Var}{Var}
\DeclareMathOperator{\logit}{logit}

\newcommand{\E}{\mathbb E}
\newcommand{\R}{\mathbb R}
\newcommand{\Nzero}{\mathbb N_0}
\newcommand{\ind}{\mathbf 1}
\newcommand{\xb}{\bm x}
\newcommand{\Xb}{\bm X}
\newcommand{\Ab}{\bm A}
\newcommand{\cb}{\bm c}
\newcommand{\ub}{\bm u}
\newcommand{\qb}{\bm q}
\newcommand{\db}{\bm d}
\newcommand{\wb}{\bm w}
\newcommand{\betab}{\bm\beta}
\newcommand{\gammab}{\bm\gamma}
\newcommand{\mub}{\bm\mu}
\newcommand{\Sigmab}{\bm\Sigma}
\newcommand{\etab}{\bm\eta}
\newcommand{\bb}{\bm b}
\newcommand{\thetab}{\bm\theta}
\newcommand{\calX}{\mathcal X}
\newcommand{\calW}{\mathcal W}
\newcommand{\calS}{\mathcal S}

\newif\ifshowdraftnotes
\showdraftnotestrue

\title{Cluster-weighted modeling for count data with excess zeros and mixed-type covariates}
\author[1]{Jongik Park}
\author[2]{Geoffrey J. McLachlan}
\author[1]{Sangkon Oh\thanks{Corresponding author. Email: \texttt{ohsangkon@pknu.ac.kr}.}}
\affil[1]{Department of Statistics and Data Science, Pukyong National University}
\affil[2]{School of Mathematics and Physics, University of Queensland}
\date{}

\begin{document}
\maketitle

\begin{abstract}
Count data may exhibit excess zeros, within-group overdispersion, and heterogeneous relationships between the response and its covariates.
Conventional zero-inflated regression models typically impose a single count-regression structure, whereas Poisson count components cannot accommodate residual within-component overdispersion. We propose a zero-inflated negative binomial cluster-weighted model that combines
a structural-zero component with multiple negative binomial regression components and jointly models the response and mixed-type covariates.
We establish sufficient conditions for identifiability of the proposed model and its Poisson counterpart and develop an expectation-maximization algorithm for estimation. Numerical studies demonstrate improved regression coefficient estimation and clustering performance relative to the Poisson counterpart under within-component overdispersion.
\end{abstract}


\section{Introduction}\label{sec:intro}

Count data arise in applications such as healthcare utilization, road safety, and insurance claims. Their analysis is often complicated by excess zeros, overdispersion, and latent heterogeneity. Poisson
regression provides a standard starting point, but its assumption of equal conditional mean and variance may be restrictive. Zero-inflated Poisson (ZIP) regression accommodates excess zeros by combining a point
mass at zero with a Poisson count component, allowing observed zeros to arise from either a structural-zero state or the count-generating process \citep{Lambert1992}. Zero-inflated negative binomial (ZINB) regression replaces the Poisson component with a negative binomial distribution and additionally accommodates overdispersion within the
count-generating population \citep{Greene1994}.

Conventional ZIP and ZINB regressions nevertheless typically specify a single regression relationship for the count-generating population. Latent subgroups may instead have different regression relationships, residual dispersion levels, and covariate distributions. Finite mixture models provide a general framework for representing such latent heterogeneity \citep{McLachlanPeel2000}. Finite mixtures of regressions allow heterogeneous response--covariate relationships through component-specific regression parameters \citep{DeSarboCron1988}, while mixtures of experts further allow the mixing proportions to depend on covariates \citep{Jacobs1991}. Recent extensions of these models have introduced semiparametric error distributions and flexible regression functions to reduce within-group model misspecification \citep{OhSeo2024,HwangSeoOh2025,OhLachosSeo2026}. However, these approaches are formulated conditionally on the covariates and do not explicitly model their distribution. Consequently, heterogeneity in the covariate distribution itself is not directly used to characterize the latent groups.

Cluster-weighted models (CWMs) address this limitation by explicitly modeling the joint distribution of the response and covariates \citep{Gershenfeld1997}. Each component combines a covariate distribution with a conditional response distribution, allowing heterogeneity in both the covariate profiles and the response--covariate relationships to contribute to component allocation. Generalized linear CWMs
accommodate non-Gaussian responses and mixed-type covariates
\citep{Ingrassia2015,Mazza2018}. More recent developments have extended
the CWM framework to a broader range of data structures and distributional
assumptions. These include matrix-normal CWMs for matrix-variate data
\citep{Tomarchio2021}, multivariate and matrix-variate CWMs based on
skewed distributions \citep{Gallaugher2022b, Gallaugher2022a},
functional CWMs \citep{SaeidiAminghafari2026}, and fully Bayesian
Gaussian CWMs \citep{PapastamoulisPerrakis2026}. The distinction
between mixture components used to approximate a distribution and
substantively meaningful clusters is important in this setting.
\citet{Baudry2010} address this distinction through component merging
in model-based clustering. Within the CWM framework, \citet{OhSeo2023}
model each cluster using multiple Gaussian components to improve
distributional flexibility while retaining interpretable regression
groups.

Zero-inflated extensions of CWMs have also been investigated.
\citet{Pocuca2020} develop zero-inflated generalized CWMs for insurance
claims, with a ZIP regression specified within each mixture component.
Related Poisson formulations are considered by \citet{Olobatuyi2022}.
However, within the count-generating state, the Poisson component still
imposes equidispersion conditional on the covariates and latent component.
When residual overdispersion remains within a component, additional
Poisson components may be required to approximate that variation,
potentially obscuring the distinction between distributional
approximation and substantively distinct regression groups. This
motivates a formulation that explicitly accommodates structural zeros,
heterogeneous regression relationships, and residual overdispersion
within the count components.

We develop a zero-inflated negative binomial cluster-weighted model  consisting of one structural-zero component and multiple negative binomial regression components. Unlike existing zero-inflated CWM formulations that specify a separate zero-inflation mechanism within each component, the proposed model represents the structural-zero
population by a single component with its own covariate distribution. The remaining count components have component-specific regression coefficients, dispersion parameters, and covariate distributions. Continuous, count, and categorical covariates are accommodated within a common joint-model framework, and the conditional structural-zero
probability is induced by the component-specific covariate distributions rather than specified through a separate logit regression. We also consider the corresponding zero-inflated Poisson cluster-weighted model.
We further establish sufficient conditions for identifiability of both models, building on classical finite-mixture results \citep{Teicher1963,YakowitzSpragins1968} and regression-mixture identifiability \citep{Hennig2000}. Under the stated conditions, the joint-model parameters are identifiable up to a common component permutation without requiring distinct Gaussian covariate distributions.
The results also cover settings without continuous covariates.
Numerical studies demonstrate the practical performance of the proposed models.

The remainder of the paper is organized as follows.
Section~\ref{sec:background} reviews zero-inflated regression and cluster-weighted modeling. Section~\ref{sec:proposed} presents the
models, identifiability results, estimation algorithm, and
model-selection criteria. Section~\ref{sec:sim} reports the simulation studies, and Section~\ref{sec:real} presents the real data analyses. Section~\ref{sec:discussion} concludes with a discussion of the findings, limitations, and directions for future research.

\section{Background}\label{sec:background}

\subsection{Zero-inflated regression}\label{sec:zi-regression}

Let $Y\in\Nzero=\{0,1,2,\ldots\}$ be a count response. We denote the Poisson probability mass function by
\begin{equation}\label{eq:poisson}
 f_{\mathrm P}(y;\lambda)
 =\frac{e^{-\lambda}\lambda^y}{y!},
 \qquad y\in\Nzero,\quad \lambda>0.
\end{equation}
For the negative binomial (NB) distribution, we use the
mean--dispersion parameterization
\begin{equation}\label{eq:nb}
 f_{\mathrm{NB}}(y;\lambda,\alpha)
 =\frac{\Gamma(y+\alpha^{-1})}
        {\Gamma(\alpha^{-1})y!}
  \left(\frac{1}{1+\alpha\lambda}\right)^{\alpha^{-1}}
  \left(\frac{\alpha\lambda}{1+\alpha\lambda}\right)^y,
 \qquad y\in\Nzero,\quad \lambda>0,\quad\alpha>0.
\end{equation}
The mean and variance of the NB distribution are
$\lambda$ and $\lambda+\alpha\lambda^2$, respectively. For each
fixed $\lambda>0$ and $y\in\Nzero$,
\[
 \lim_{\alpha\downarrow0}
 f_{\mathrm{NB}}(y;\lambda,\alpha)
 =f_{\mathrm P}(y;\lambda).
\]
Thus, the Poisson distribution is a boundary limit of the negative
binomial family, while~\eqref{eq:nb} itself is defined for
$\alpha>0$.

Let $\delta_0(y)=\ind(y=0)$ denote the probability mass function of a point mass at zero. A conventional zero-inflated regression combines this structural-zero component with a count component.
Using the same covariate vector $\xb$ for the zero-inflation and count parts, its conditional probability mass function can be written as
\begin{equation*}\label{eq:zi-regression}
 p_\Theta(y\mid\xb)
 =
 \begin{cases}
 \zeta(\xb)\delta_0(y)
 +\{1-\zeta(\xb)\}
   f_{\mathrm P}\{y;\lambda(\xb)\},
 & \text{if ZIP},\\[4pt]
 \zeta(\xb)\delta_0(y)
 +\{1-\zeta(\xb)\}
   f_{\mathrm{NB}}\{y;\lambda(\xb),\alpha\},
 & \text{if ZINB}.
 \end{cases}
\end{equation*}
Here, $\zeta(\xb)$ is the structural-zero probability and
$\lambda(\xb)$ is the mean of the count-generating component.
Common link functions are
\[
 \log\lambda(\xb)=\tilde\xb^\top\betab,
 \qquad
 \logit\{\zeta(\xb)\}
 =\tilde\xb^\top\gammab,
\]
where $\tilde\xb$ includes an intercept
\citep{Lambert1992,Greene1994}.

The conditional mean and variance of the ZINB regression are
\begin{equation}\label{eq:zinb-moments}
 \begin{aligned}
  \E(Y\mid\xb)
  &=
  \{1-\zeta(\xb)\}\lambda(\xb),\\
  \Var(Y\mid\xb)
  &=
  \{1-\zeta(\xb)\}\lambda(\xb)
  \left[
    1+\{\zeta(\xb)+\alpha\}\lambda(\xb)
  \right].
 \end{aligned}
\end{equation}
The corresponding ZIP moments are obtained by setting
$\alpha=0$ in~\eqref{eq:zinb-moments}. Both models distinguish structural zeros from zeros generated by the count component, but assume a single regression relationship for the count-generating population.
This assumption can be restrictive when latent subgroups exhibit heterogeneous regression relationships, residual dispersion levels, or covariate distributions, motivating mixture-based approaches.

\subsection{Cluster-weighted modeling}\label{sec:cwm-background}

Let $Z\in\{1,\ldots,K\}$ denote an unobserved component label with $\Pr(Z=k)=\pi_k$. A cluster-weighted model specifies the joint distribution of the response and covariates as \begin{equation}\label{eq:cwm-background}
 p_\Theta(y,\xb)
 =\sum_{k=1}^{K}
   \pi_k g_k(\xb;\etab_k)
   f_k(y\mid\xb;\thetab_k),
 \qquad
 \pi_k>0,\quad
 \sum_{k=1}^{K}\pi_k=1.
\end{equation}
Here, $g_k(\xb;\etab_k)$ describes the covariate distribution
within component $k$, and $f_k(y\mid\xb;\thetab_k)$ is the
corresponding conditional response probability mass function
\citep{Gershenfeld1997,Ingrassia2015}. The parameters $\etab_k$
and $\thetab_k$ govern the covariate and conditional response
distributions, respectively.

The marginal covariate distribution and conditional component
weights are
\begin{equation}\label{eq:weights}
 \begin{aligned}
  p_X(\xb)
  &=\sum_{j=1}^{K}\pi_jg_j(\xb;\etab_j),\\
  \omega_k(\xb)
  &=\Pr(Z=k\mid\Xb=\xb)
    =\frac{\pi_kg_k(\xb;\etab_k)}{p_X(\xb)}.
 \end{aligned}
\end{equation}
The dependence of $p_X$ and $\omega_k$ on $\Theta$ is suppressed throughout for notational simplicity. The weights $\pi_k$ are unconditional component probabilities, whereas $\omega_k(\xb)$ conditions on the covariates but not on the response. Consequently,
\[
 p_\Theta(y\mid\xb)
 =\sum_{k=1}^{K}
   \omega_k(\xb)f_k(y\mid\xb;\thetab_k).
\]

For mixed-type covariates, different distributional families can be assigned to different covariate blocks
\citep{Ingrassia2015,Mazza2018}. In the present work, the continuous block follows a multivariate Gaussian distribution, each count covariate follows a Poisson distribution, and each categorical variable follows a categorical distribution. Conditional on $Z$,
the continuous block, the individual count covariates, and the categorical variables are mutually independent. Dependence among the continuous covariates is accommodated by an unrestricted within-component covariance matrix. 

\section{Zero-inflated cluster-weighted modeling}
\label{sec:proposed}

\subsection{Model formulation}\label{sec:model}

Let $Z\in\{1,\ldots,K\}$ denote the latent component label,
with $\Pr(Z=k)=\pi_k$, where $\pi_k>0$ and $\sum_{k=1}^{K}\pi_k=1$. We reserve component $k=1$ for the
structural-zero population and components $k=2,\ldots,K$ for
the count-generating populations. Partition the covariate vector as
\[
 \xb=(\qb^\top,\db^\top,\wb^\top)^\top,
\]
where
\[
 \qb=(q_1,\ldots,q_{p_Q})^\top\in\R^{p_Q},
 \qquad
 \db=(d_1,\ldots,d_{p_D})^\top\in\Nzero^{p_D},
\]
and
\[
 \wb=(w_1,\ldots,w_{p_W})^\top\in\calW,
 \qquad
 \calW
 :=\prod_{\ell=1}^{p_W}\{1,\ldots,r_\ell\}.
\]
Here, $\qb$, $\db$, and $\wb$ contain the continuous,
count-valued, and categorical covariates, respectively, and
$r_\ell$ denotes the number of categories of categorical
covariate $\ell$. Any of the three covariate blocks may be absent.

Conditional on $Z=k$, the continuous block, individual count
covariates, and categorical covariates are assumed to be mutually independent, while dependence among the continuous covariates is allowed through an unrestricted covariance matrix. Specifically, the covariate distribution within component $k$ is
\begin{equation}\label{eq:covariates}
 g_k(\xb;\etab_k)
 =\phi_{p_Q}(\qb;\mub_k,\Sigmab_k)
   \prod_{j=1}^{p_D}f_{\mathrm P}(d_j;\nu_{kj})
   \prod_{\ell=1}^{p_W}\prod_{c=1}^{r_\ell}
       \rho_{k\ell c}^{\ind(w_\ell=c)}.
\end{equation}
Here, $\phi_{p_Q}(\,\cdot\,;\mub_k,\Sigmab_k)$ is the
$p_Q$-variate Gaussian density with mean $\mub_k$ and positive definite covariance matrix $\Sigmab_k$,
\[
 D_j\mid Z=k\sim\operatorname{Poisson}(\nu_{kj}),
 \qquad \nu_{kj}>0,
\]
and
\[
 \rho_{k\ell c}
 =\Pr(W_\ell=c\mid Z=k),
 \qquad
 \rho_{k\ell c}>0,
 \qquad
 \sum_{c=1}^{r_\ell}\rho_{k\ell c}=1.
\]
Here, $D_j$ and $W_\ell$ denote the $j$th count covariate and
the $\ell$th categorical covariate, respectively. Thus,
$\nu_{kj}$ denotes the component-specific mean of count
covariate $j$. The Poisson mass function in~\eqref{eq:poisson}
is used for the count covariates and, later, for the response
components of ZIPCWM.

The covariate distribution $g_k$ is defined on the common support
\[
 \calX
 =\R^{p_Q}\times\Nzero^{p_D}\times\calW,
\]
with respect to the corresponding product of Lebesgue and counting
measures. Factors corresponding to absent covariate blocks are interpreted as one. In particular, if $p_W=0$, $\calW$ is
interpreted as a singleton containing the empty categorical
configuration. The parameter collection $\etab_k$ contains
$\mub_k$, $\Sigmab_k$, the count-covariate means $\nu_{kj}$,
and the category probabilities $\rho_{k\ell c}$ for the blocks that are present.

For each categorical covariate, choose one category as the
reference level, and let $\tilde\wb$ denote the resulting vector of reference-category dummy variables. Define the augmented regression vector by
\[
 \tilde\xb
 =(1,\qb^\top,\db^\top,\tilde\wb^\top)^\top
 \in\R^{p_x},
 \qquad
 p_x
 =1+p_Q+p_D+\sum_{\ell=1}^{p_W}(r_\ell-1).
\]
For each count component $k=2,\ldots,K$, define
\[
 \lambda_k(\xb)
 =\exp(\tilde\xb^\top\betab_k),
 \qquad
 \betab_k=(\beta_{k0},\ldots,\beta_{k,p_x-1})^\top
 \in\R^{p_x}.
\]
Here, $\lambda_k(\xb)$ is the conditional mean of the response
within count component $k$. The first subscript in $\beta_{kj}$
indexes the component, and $j=0$ denotes the intercept.

The proposed zero-inflated negative binomial cluster-weighted
model (ZINBCWM) is
\begin{equation}\label{eq:zinbcwm}
 p_\Theta(y,\xb)
 =\pi_1g_1(\xb;\etab_1)\delta_0(y)
  +\sum_{k=2}^{K}
    \pi_kg_k(\xb;\etab_k)
    f_{\mathrm{NB}}\{y;\lambda_k(\xb),\alpha_k\},
\end{equation}
where $\alpha_k>0$ is the dispersion parameter of count
component $k$. In the notation of~\eqref{eq:cwm-background},
\[
 f_1(y\mid\xb)=\delta_0(y),
\]
whereas, for $k=2,\ldots,K$,
\[
 f_k(y\mid\xb;\thetab_k)
 =f_{\mathrm{NB}}\{y;\lambda_k(\xb),\alpha_k\},
 \qquad
 \thetab_k=(\betab_k^\top,\alpha_k)^\top.
\]
Thus, $K$ denotes the total number of components: one
structural-zero component and $K-1$ count components. The
structural-zero component has its own covariate distribution
but no response-regression coefficients or dispersion parameter.
The parameter collection is
\[
 \Theta
 =\left(
   K,\{\pi_k,\etab_k\}_{k=1}^{K},
   \{\betab_k,\alpha_k\}_{k=2}^{K}
  \right).
\]

The corresponding zero-inflated Poisson cluster-weighted
model (ZIPCWM) uses the same covariate specification and
replaces the negative binomial response distribution in each
count component by a Poisson distribution:
\begin{equation}\label{eq:zipcwm}
 p_\Theta(y,\xb)
 =\pi_1g_1(\xb;\etab_1)\delta_0(y)
  +\sum_{k=2}^{K}
    \pi_kg_k(\xb;\etab_k)
    f_{\mathrm P}\{y;\lambda_k(\xb)\}.
\end{equation}
For ZIPCWM, $\thetab_k=\betab_k$ for $k=2,\ldots,K$, and the
dispersion parameters are omitted from $\Theta$. Both models
define normalized joint distributions of $(Y,\Xb)$. In
particular, $g_1(\xb;\etab_1)$ describes the covariate
distribution of the structural-zero population.

Using the marginal covariate distribution and conditional
component weights in~\eqref{eq:weights}, the conditional
response distribution under ZINBCWM is
\begin{equation}\label{eq:conditional}
 p_\Theta(y\mid\xb)
 =\omega_1(\xb)\delta_0(y)
  +\sum_{k=2}^{K}
    \omega_k(\xb)
    f_{\mathrm{NB}}\{y;\lambda_k(\xb),\alpha_k\}.
\end{equation}
Hence, the conditional probability of belonging to the
structural-zero component is $\omega_1(\xb)$, which is determined by the mixing proportions and component-specific covariate distributions.

The conditional mean and variance under ZINBCWM follow from
the laws of total expectation and total variance:
\begin{align}
 m(\xb)
 &:=\E(Y\mid\xb)
   =\sum_{k=2}^{K}\omega_k(\xb)\lambda_k(\xb),
 \label{eq:mean}\\
 \Var(Y\mid\xb)
 &=\sum_{k=2}^{K}
   \omega_k(\xb)
   \{\lambda_k(\xb)+\alpha_k\lambda_k(\xb)^2\}
 \notag\\
 &\quad
   +\omega_1(\xb)m(\xb)^2
   +\sum_{k=2}^{K}
    \omega_k(\xb)
    \{\lambda_k(\xb)-m(\xb)\}^2.
 \label{eq:variance}
\end{align}
The first sum in~\eqref{eq:variance} represents
within-component variation. The remaining terms represent
variation among the component-specific conditional means,
including the zero mean of the structural-zero component.
The conditional mean of ZIPCWM has the same form as
\eqref{eq:mean}, and its conditional variance is obtained by
setting $\alpha_k=0$, $k=2,\ldots,K$, in
\eqref{eq:variance}.

\subsection{Identifiability}\label{sec:identifiability}

We first establish a distribution-level identifiability result for the response components. The argument builds on the transform-ratio criterion of \citet{Teicher1963} and the probability-generating-function approach for negative binomial mixtures in \citet{YakowitzSpragins1968}. We extend these arguments to the combined family that also includes Poisson distributions and a point mass at zero.

\begin{lemma}
\label{lem:count}
Let
\[
 \mathcal F
 =
 \{\delta_0\}
 \cup
 \{f_{\mathrm{NB}}(\cdot;\lambda,\alpha):
      \lambda>0,\ \alpha>0\}
 \cup
 \{f_{\mathrm P}(\cdot;\lambda):\lambda>0\}.
\]
Finite mixtures of distinct members of $\mathcal F$ with
strictly positive mixing weights are identifiable up to a
permutation of their components.
\end{lemma}

\begin{proof}
A proof is given in Appendix~A.
\end{proof}

We next introduce notation for establishing identifiability of the regression components. In particular, it is useful to separate the continuous and count covariates from the categorical covariates, because, for a fixed categorical configuration, the component log mean is an affine function of the remaining covariates.

Partition the regression coefficients as
\[
 \betab_k
 =
 \bigl(
   \beta_{k0},
   \betab_{k,Q}^\top,
   \betab_{k,D}^\top,
   \betab_{k,W}^\top
 \bigr)^\top,
 \qquad
 \bb_k
 =
 (\betab_{k,Q}^\top,\betab_{k,D}^\top)^\top,
\]
and put
\[
 \ub=(\qb^\top,\db^\top)^\top.
\]
Thus, when convenient, a function of $\xb$ may equivalently be
written as a function of $(\ub,\wb)$. Define the component log
mean by
\begin{equation}\label{eq:linear-predictor}
 \xi_k(\ub,\wb)
 :=\log\lambda_k(\ub,\wb)
 =a_k(\wb)+\ub^\top\bb_k,
 \qquad
 a_k(\wb)
 =\beta_{k0}+\tilde\wb^\top\betab_{k,W}.
\end{equation}
For a fixed categorical configuration $\wb$,
$\xi_k(\ub,\wb)$ is therefore an affine function of $\ub$.
Throughout, $\xi_k$ denotes the log mean of count component $k$, whereas $m(\xb)$ in~\eqref{eq:mean} denotes the overall
conditional mean.

Throughout the following results, the covariate dimensions,
category sets, and reference-category coding are fixed. We use the following common assumptions for the cluster-weighted models.

\Needspace{5\baselineskip}
\medskip
\noindent\textbf{Assumptions.}
\begin{enumerate}[
  label=(A\arabic*),
  leftmargin=3.8em,
  itemsep=5pt
]
\item\label{ass:regular}
The number of components is finite, $K\ge2$, with
$\pi_k>0$ and
\[
 \sum_{k=1}^{K}\pi_k=1.
\]

\item\label{ass:cov}
The component covariate distributions are given
by~\eqref{eq:covariates}, with $\Sigmab_k\succ0$,
$\nu_{kj}>0$, and $\rho_{k\ell c}>0$, where
\[
 \sum_{c=1}^{r_\ell}\rho_{k\ell c}=1.
\]
Conditions corresponding to absent covariate blocks are omitted.
\end{enumerate}

It follows from (A2) that all component covariate distributions are strictly positive on the common support
\[
 \calX=\R^{p_Q}\times\Nzero^{p_D}\times\calW.
\]
No separation between the component covariate distributions is required. Conditions distinguishing the count components are stated separately in the individual identifiability results below.

The following technical lemma establishes a polynomial uniqueness property on the mixed continuous--integer support. This property will be used to obtain a common correspondence between count components and to recover the component covariate distributions over the entire support.

\begin{lemma}
\label{lem:polynomial}
Let $\calS=\R^{p_Q}\times\Nzero^{p_D}$. A real polynomial
$P(\ub)$ that vanishes at every point of $\calS$ is the
zero polynomial. Moreover, if $H(\ub)$ is a nonzero real
polynomial and $P(\ub)=0$ for every $\ub\in\calS$ with
$H(\ub)\ne0$, then $P$ is the zero polynomial. If
$p_Q+p_D=0$, interpret $\calS$ as a singleton and the
polynomial ring as the real constants.
\end{lemma}

\begin{proof}
For a univariate polynomial, vanishing on $\R$ or on
$\Nzero$ implies infinitely many distinct roots and therefore
the polynomial is identically zero. For several variables,
view the polynomial as a polynomial in the last coordinate,
with coefficients that are polynomials in the remaining
coordinates. Fixing the remaining coordinates at any point
of their product support shows that all such coefficients
vanish on that support. Repeating this argument over the
remaining coordinates proves the first assertion by induction
on the dimension. The zero-dimensional case is immediate.

For the second assertion, consider $P(\ub)H(\ub)$. If
$H(\ub)\ne0$, then $P(\ub)=0$ by assumption, and hence
$P(\ub)H(\ub)=0$. If $H(\ub)=0$, then
$P(\ub)H(\ub)=0$ automatically. Therefore,
\[
 P(\ub)H(\ub)=0
 \qquad\text{for every }\ub\in\calS.
\]
By the first assertion,
\[
 P(\ub)H(\ub)\equiv0.
\]
Since a polynomial ring over $\R$ has no zero divisors and
$H(\ub)$ is not the zero polynomial, it follows that
$P(\ub)\equiv0$.
\end{proof}

The preceding lemmas provide the two main ingredients for
establishing identifiability of the full ZINBCWM. Lemma~\ref{lem:count} identifies the response components at covariate values where the corresponding count distributions are distinct, while Lemma~\ref{lem:polynomial} allows the resulting component correspondence to be extended across the mixed continuous--integer covariate support. We now combine these results to establish identifiability of the mixing proportions, regression parameters, and component covariate distributions.

\begin{theorem}
\label{thm:ident}
Consider two ZINBCWM parameter collections $\Theta$ and $\Theta'$, with $K$ and $K'$ components, respectively, each satisfying (A1)--(A2), with strictly positive and finite dispersion parameters for all count components. Suppose that, within each parameter collection, the count-component pairs $(\betab_k,\alpha_k)$ are pairwise distinct when $p_W=0$, and that $(\bb_k,\alpha_k)$ are pairwise distinct when $p_W>0$.

If
\[
 p_\Theta(y,\xb)=p_{\Theta'}(y,\xb)
\]
for every $y\in\Nzero$ and almost every $\xb$ with respect
to the product of Lebesgue and counting measures, then $K'=K$.
Moreover, there exists a permutation $\tau$ of
$\{1,\ldots,K\}$ with $\tau(1)=1$ such that
\begin{align*}
 \pi_k&=\pi'_{\tau(k)},
 &\qquad\etab_k&=\etab'_{\tau(k)},
 &&k=1,\ldots,K,\\
 \betab_k&=\betab'_{\tau(k)},
 &\alpha_k&=\alpha'_{\tau(k)},
 &&k=2,\ldots,K.
\end{align*}
\end{theorem}

\begin{proof}
A proof is given in Appendix~B.
\end{proof}

Theorem~\ref{thm:ident} also covers the case of purely categorical covariates. When $p_Q=p_D=0$, the vector $\bb_k$ is empty, so the separation condition on $(\bb_k,\alpha_k)$ reduces to pairwise distinctness of the dispersion parameters. Thus, the ZINBCWM remains identifiable without continuous or count covariates, provided that the negative binomial count components have distinct dispersions.

The argument used in the proof of Theorem~\ref{thm:ident} also applies to the Poisson counterpart of the proposed model. In the absence of dispersion parameters, two Poisson response components coincide exactly when their conditional means coincide. Thus, the same polynomial argument can be used to exclude such collisions and to establish a common component correspondence across the covariate support. This yields the following identifiability result for the ZIPCWM.

\Needspace{6\baselineskip}
\begin{corollary}
\label{cor:poisson}
Under (A1)--(A2), consider the class of ZIPCWMs
in~\eqref{eq:zipcwm} for which the count-regression vectors
$\betab_k$, $k=2,\ldots,K$, are pairwise distinct when
$p_W=0$, and the vectors $\bb_k$, $k=2,\ldots,K$, are
pairwise distinct when $p_W>0$. This class is identifiable,
including the number of components, up to a permutation of the count components.
\end{corollary}

\begin{proof}
Let $k,\ell\in\{2,\ldots,K\}$ denote two distinct count
components. By Lemma~\ref{lem:count}, at any covariate value
$\xb$ for which the conditional means $\lambda_k(\xb)$, $k=2,\ldots,K$, are pairwise distinct, the corresponding Poisson distributions and their conditional weights are identifiable up to a permutation of the count components.

For each pair $2\leq k<\ell\leq K$, a collision between the
$k$th and $\ell$th Poisson components occurs only when
\[
\lambda_k(\xb)=\lambda_\ell(\xb),
\]
or, equivalently, when their affine log-mean functions coincide.
The collision polynomial obtained from the pairwise differences of these affine functions is therefore nonzero under the stated pairwise-distinctness condition. The same polynomial argument as in the proof of Theorem~\ref{thm:ident}, together with
Lemma~\ref{lem:polynomial}, then extends the pointwise component correspondence to the entire covariate support and identifies the regression functions up to a common permutation.

When $p_W>0$, pairwise distinctness of the vectors $\bb_k$,
$k=2,\ldots,K$, ensures that the same component labeling can be used across categorical configurations. The reference-category argument then identifies the remaining regression coefficients.
Finally, the log-ratio argument used in the proof of
Theorem~\ref{thm:ident} identifies $\pi_k g_k(\xb;\etab_k)$ for each component. Integration and normalization identify the mixing proportions and covariate parameters. Hence, the ZIPCWM, including the number of components, is identifiable up to a permutation of the count components.
\end{proof}

\subsection{Estimation}\label{sec:em}

For a fixed candidate $K$, let $(y_i,\xb_i)$, $i=1,\ldots,n$, be independent observations, and write
$\lambda_{ik}=\lambda_k(\xb_i)=\exp(\tilde\xb_i^\top\betab_k)$.
The observed log-likelihood of ZINBCWM is
\begin{equation*}\label{eq:loglik}
 \ell(\Theta)
 =\sum_{i=1}^{n}\log\left[
   \pi_1g_1(\xb_i;\etab_1)\delta_0(y_i)
   +\sum_{k=2}^{K}\pi_kg_k(\xb_i;\etab_k)
     f_{\mathrm{NB}}(y_i;\lambda_{ik},\alpha_k)\right].
\end{equation*}
For ZIPCWM, $f_{\mathrm{NB}}(y_i;\lambda_{ik},\alpha_k)$ is replaced by $f_{\mathrm P}(y_i;\lambda_{ik})$.

We use an EM-based algorithm \citep{Dempster1977}. Let $Z_i$ be the component label of observation $i$, and define
$Z_{ik}=\ind(Z_i=k)$. At iteration $t$, write
$\lambda_{ik}^{(t)}=\exp(\tilde\xb_i^\top\betab_k^{(t)})$.
The E-step computes
\[
 z_{ik}^{(t)}
 =\E(Z_{ik}\mid y_i,\xb_i;\Theta^{(t)})
 =\Pr(Z_i=k\mid Y_i=y_i,\Xb_i=\xb_i;\Theta^{(t)}).
\]
For ZINBCWM, these posterior component probabilities are
\begin{equation*}\label{eq:estep}
 \begin{aligned}
  z_{i1}^{(t)}
  &=\frac{\pi_1^{(t)}g_1(\xb_i;\etab_1^{(t)})\delta_0(y_i)}
          {p_{\Theta^{(t)}}(y_i,\xb_i)},\\[3pt]
  z_{ik}^{(t)}
  &=\frac{\pi_k^{(t)}g_k(\xb_i;\etab_k^{(t)})
         f_{\mathrm{NB}}(y_i;\lambda_{ik}^{(t)},\alpha_k^{(t)})}
          {p_{\Theta^{(t)}}(y_i,\xb_i)},
    \qquad k=2,\ldots,K.
 \end{aligned}
\end{equation*}
In particular, $y_i>0$ implies $z_{i1}^{(t)}=0$.
The posterior probabilities $z_{ik}^{(t)}$ condition on both the response and covariates, unlike the covariate-conditional weights $\omega_k(\xb)$ in~\eqref{eq:weights}. The ZIPCWM E-step replaces the negative binomial mass function by $f_{\mathrm P}(y_i;\lambda_{ik}^{(t)})$ in both the numerator and the corresponding denominator.

Because $Z_{i1}=1$ is compatible only with $y_i=0$, the contribution $Z_{i1}\log\delta_0(y_i)$ is zero on the support of the complete data.
Up to constants, the expected complete-data log-likelihood is
\begin{equation*}\label{eq:qfunction}
\begin{aligned}
 \mathcal Q(\Theta\mid\Theta^{(t)})
 ={}&\sum_{i=1}^{n}\sum_{k=1}^{K}z_{ik}^{(t)}
           \{\log\pi_k+\log g_k(\xb_i;\etab_k)\}\\
 &+\sum_{i=1}^{n}\sum_{k=2}^{K}z_{ik}^{(t)}
           \log f_{\mathrm{NB}}(y_i;\lambda_{ik},\alpha_k).
\end{aligned}
\end{equation*}
For observations with $y_i=0$, membership in the structural-zero component remains latent rather than being deterministically assigned, and the posterior probabilities are retained throughout the updates.

For the covariate blocks that are present, the interior unconstrained updates are
\begin{align*}
 \pi_k^{(t+1)} 
 =\frac{\sum_{i=1}^{n}z_{ik}^{(t)}}{n}, \qquad 
  \nu_{kj}^{(t+1)} 
 =\frac{\sum_{i=1}^{n}z_{ik}^{(t)}d_{ij}}
         {\sum_{i=1}^{n}z_{ik}^{(t)}},  \qquad
 \rho_{k\ell c}^{(t+1)} 
 =\frac{\sum_{i=1}^{n}z_{ik}^{(t)}
   \ind(w_{i\ell}=c)}
   {\sum_{i=1}^{n}z_{ik}^{(t)}}, 
\\
 \mub_k^{(t+1)}
 =\frac{\sum_{i=1}^{n}z_{ik}^{(t)}\qb_i}
         {\sum_{i=1}^{n}z_{ik}^{(t)}}, \qquad
 \Sigmab_k^{(t+1)}
 =\frac{\sum_{i=1}^{n}z_{ik}^{(t)}
   (\qb_i-\mub_k^{(t+1)})
   (\qb_i-\mub_k^{(t+1)})^\top}
   {\sum_{i=1}^{n}z_{ik}^{(t)}}.
\end{align*}

For each count component, with the posterior probabilities
$z_{ik}^{(t)}$ fixed, maximize
\begin{equation*}\label{eq:count-mstep}
 \mathcal Q_{Y,k}(\betab_k,\alpha_k)
 =\sum_{i=1}^{n}z_{ik}^{(t)}
   \log f_{\mathrm{NB}}(y_i;\lambda_{ik},\alpha_k)
\end{equation*}
over $\betab_k$ and $\alpha_k>0$. The weighted regression score is
\[
 \frac{\partial\mathcal Q_{Y,k}}{\partial\betab_k}
 =\sum_{i=1}^{n}z_{ik}^{(t)}
   \frac{y_i-\lambda_{ik}}
        {1+\alpha_k\lambda_{ik}}\tilde\xb_i.
\]
The dispersion score is
\begin{equation*}\label{eq:alpha-score}
 \frac{\partial\mathcal Q_{Y,k}}{\partial\alpha_k}
 =\sum_{i=1}^{n}z_{ik}^{(t)}\left[
 \frac{\log(1+\alpha_k\lambda_{ik})
       -\psi(y_i+\alpha_k^{-1})
       +\psi(\alpha_k^{-1})}
      {\alpha_k^2}
 +\frac{y_i-\lambda_{ik}}
       {\alpha_k(1+\alpha_k\lambda_{ik})}
 \right],
\end{equation*}
where $\psi$ is the digamma function. We update $\betab_k$ and
$\alpha_k$ jointly by numerical maximization of
$\mathcal Q_{Y,k}$.

For ZIPCWM, the corresponding objective is the weighted Poisson
log-likelihood
\[
 \mathcal Q_{Y,k}(\betab_k)
 =\sum_{i=1}^{n}z_{ik}^{(t)}
   \log f_{\mathrm P}(y_i;\lambda_{ik}),
\]
and there is no dispersion update.

\section{Simulation study}
\label{sec:sim}

In this section, we conduct simulation studies to evaluate the performance of the proposed ZINBCWM in comparison with the ZIPCWM. We mainly focus on regression parameter estimation and clustering performance. For the regression parameter estimation when the true number of components is two, the conventional ZIP and ZINB models are also considered for comparison.

For all simulation studies, the true number of components is
$K_{\mathrm{true}}\in\{2,3\}$. Component 1 represents the
structural-zero component, and the component probabilities are given by
\[
 \Pr(Z=1)=\pi_1,
 \qquad
 \Pr(Z=k)=\frac{1-\pi_1}{K_{\mathrm{true}}-1},
 \quad k=2,\ldots,K_{\mathrm{true}},
\]
where $\pi_1\in\{0.2,0.5\}$. When $K_{\mathrm{true}}=2$, only components 1 and 2 are used.

We generate two continuous covariates, one count covariate, and one binary categorical covariate. Thus, $p_Q=2$, $p_D=1$, $p_W=1$, and $r_1=2$. Conditional on component membership, the covariates are generated from
\[
 g_k(\xb;\etab_k)
 =
 \phi_2(\qb;\mub_k,\Sigmab)
 f_{\mathrm P}(d_1;\nu_{k1})
 \rho_{k11}^{\ind(w_1=1)}
 \rho_{k12}^{\ind(w_1=2)},
\]
where
\[
 \mub_1=(0,0)^\top,\qquad
 \mub_2=(1,-1)^\top,\qquad
 \mub_3=(-1,1)^\top,
\]
\[
 \Sigmab=
 \begin{pmatrix}
 0.5&0.1\\
 0.1&0.5
 \end{pmatrix},
 \qquad
 (\nu_{11},\nu_{21},\nu_{31})=(2,4,6),
\]
and
\[
 (\rho_{112},\rho_{212},\rho_{312})
 =(0.4,0.6,0.4),
 \qquad
 \rho_{k11}=1-\rho_{k12}.
\]
The first category is used as the reference category.

For the structural-zero component, we set $Y=0$. For the count components $k=2,\ldots,K_{\mathrm{true}}$, the conditional mean is specified as
\[
 \log\lambda_k(\xb)
 =
 \beta_{k0}
 +\beta_{k1}q_1
 +\beta_{k2}q_2
 +\beta_{k3}d_1
 +\beta_{k4}\ind(w_1=2),
\]
where
\[
 \betab_2=(0,0.2,-0.2,0.2,-0.2)^\top,
 \qquad
 \betab_3=(0.5,-0.2,0.2,0.2,0.2)^\top.
\]
We consider the following four cases according to the conditional
distribution of the count components:
\begin{itemize}
\item \textbf{Case I:}
$Y\mid(\xb,Z=k)\sim
 \mathrm{Poisson}\{\lambda_k(\xb)\}$;

\item \textbf{Case II:}
$Y\mid(\xb,Z=k)\sim
 \mathrm{NB}\{\lambda_k(\xb),0.1\}$;

\item \textbf{Case III:}
$Y\mid(\xb,Z=k)\sim
 \mathrm{NB}\{\lambda_k(\xb),1\}$;

\item \textbf{Case IV:}
$Y\mid(\xb,Z=k)\sim
 \mathrm{NB}\{\lambda_k(\xb),2\}$.
\end{itemize}
Thus, Case I corresponds to the absence of overdispersion,
whereas Cases II--IV represent increasing levels of
overdispersion. We consider sample sizes
$n\in\{500,1000\}$ and generate $B=200$ replicated samples for each setting.

\subsection{Simulation 1: Regression parameter estimation}
\label{sec:sim-estimation}

In the first simulation study, we investigate the estimation
performance of the regression coefficients. For each simulated sample, ZIPCWM and ZINBCWM are fitted using the true number of components, $K=K_{\mathrm{true}}$. When
$K_{\mathrm{true}}=2$, ZIP and ZINB are also considered because only one count component is present. When
$K_{\mathrm{true}}=3$, ZIP and ZINB are excluded because they
contain only one count-regression equation.

Let $p_\beta=5(K_{\mathrm{true}}-1)$ denote the total number of regression coefficients for the count components. After matching the component labels, the bias and mean squared error (MSE) are computed as
\[
 \mathrm{Bias}
 =
 \frac{1}{p_\beta}
 \sum_{j=1}^{p_\beta}
 \left(
 \frac{1}{B}\sum_{b=1}^{B}\widehat\beta_j^{(b)}
 -\beta_j
 \right),
\]
and
\[
 \mathrm{MSE}
 =
 \frac{1}{p_\beta}
 \sum_{j=1}^{p_\beta}
 \frac{1}{B}\sum_{b=1}^{B}
 \left(
 \widehat\beta_j^{(b)}-\beta_j
 \right)^2,
\]
where $\widehat\beta_j^{(b)}$ denotes the estimate of the
$j$th regression coefficient from the $b$th replicated sample.
Tables~\ref{tab:sim-coef-k2} and~\ref{tab:sim-coef-k3}
summarize the results using MSE with Bias in parentheses.

\begin{table}[tbp]
\centering
\caption{Regression coefficient estimation performance:
MSE (Bias), $K_{\mathrm{true}}=2$.}
\label{tab:sim-coef-k2}
\resizebox{\textwidth}{!}{%
\begin{tabular}{lllcccc}
\toprule
Case & $\pi_1$ & $n$ & ZIP & ZINB & ZIPCWM & ZINBCWM \\
\midrule
\multirow{4}{*}{I}
& \multirow{2}{*}{0.2}
& 500  & 0.0035 ($-0.0011$) & 0.0036 ($-0.0016$)
       & \textbf{0.0034} ($-0.0013$) & \textbf{0.0034} ($-0.0016$) \\
& & 1000 & \textbf{0.0016} ($-0.0031$) & 0.0017 ($-0.0036$)
         & \textbf{0.0016} ($-0.0028$) & \textbf{0.0016} ($-0.0032$) \\
\cmidrule(lr){2-7}
& \multirow{2}{*}{0.5}
& 500  & \textbf{0.0065} ($-0.0013$) & \textbf{0.0065} ($-0.0022$)
       & \textbf{0.0065} ($-0.0015$) & \textbf{0.0065} ($-0.0021$) \\
& & 1000 & 0.0031 ($-0.0008$) & 0.0031 ($-0.0015$)
         & \textbf{0.0029} ($-0.0009$) & \textbf{0.0029} ($-0.0013$) \\
\midrule
\multirow{4}{*}{II}
& \multirow{2}{*}{0.2}
& 500  & 0.0066 (0.0108) & 0.0058 (0.0006)
       & 0.0063 (0.0101) & \textbf{0.0054} (0.0012) \\
& & 1000 & 0.0032 (0.0082) & 0.0025 ($-0.0025$)
         & 0.0031 (0.0065) & \textbf{0.0024} ($-0.0030$) \\
\cmidrule(lr){2-7}
& \multirow{2}{*}{0.5}
& 500  & 0.0107 (0.0128) & 0.0099 (0.0023)
       & 0.0103 (0.0108) & \textbf{0.0096} (0.0011) \\
& & 1000 & 0.0051 (0.0104) & 0.0045 ($-0.0010$)
         & 0.0049 (0.0086) & \textbf{0.0044} ($-0.0018$) \\
\midrule
\multirow{4}{*}{III}
& \multirow{2}{*}{0.2}
& 500  & 0.0570 (0.0844) & 0.0175 ($-0.0017$)
       & 0.0602 (0.0864) & \textbf{0.0162} ($-0.0014$) \\
& & 1000 & 0.0470 (0.0854) & 0.0097 ($-0.0029$)
         & 0.0503 (0.0877) & \textbf{0.0089} ($-0.0031$) \\
\cmidrule(lr){2-7}
& \multirow{2}{*}{0.5}
& 500  & 0.0743 (0.0940) & 0.0306 (0.0091)
       & 0.0802 (0.0960) & \textbf{0.0271} (0.0095) \\
& & 1000 & 0.0533 (0.0861) & 0.0137 ($-0.0004$)
         & 0.0585 (0.0887) & \textbf{0.0123} ($-0.0008$) \\
\midrule
\multirow{4}{*}{IV}
& \multirow{2}{*}{0.2}
& 500  & 0.1179 (0.1337) & 0.0303 ($-0.0070$)
       & 0.1220 (0.1365) & \textbf{0.0266} ($-0.0046$) \\
& & 1000 & 0.1071 (0.1407) & 0.0166 (0.0040)
         & 0.1116 (0.1436) & \textbf{0.0136} (0.0024) \\
\cmidrule(lr){2-7}
& \multirow{2}{*}{0.5}
& 500  & 0.1314 (0.1348) & 0.0507 (0.0038)
       & 0.1375 (0.1386) & \textbf{0.0440} (0.0034) \\
& & 1000 & 0.1184 (0.1389) & 0.0246 (0.0027)
         & 0.1251 (0.1426) & \textbf{0.0226} (0.0023) \\
\bottomrule
\end{tabular}
}
\end{table}

\begin{table}[tbp]
\centering
\caption{Regression coefficient estimation performance:
MSE (Bias), $K_{\mathrm{true}}=3$.}
\label{tab:sim-coef-k3}
\small
\begin{tabular}{lllcc}
\toprule
Case & $\pi_1$ & $n$ & ZIPCWM & ZINBCWM \\
\midrule
\multirow{4}{*}{I}
& \multirow{2}{*}{0.2}
& 500  & \textbf{0.0052} ($-0.0009$) & \textbf{0.0052} ($-0.0014$) \\
& & 1000 & \textbf{0.0026} ($-0.0025$) & 0.0027 ($-0.0028$) \\
\cmidrule(lr){2-5}
& \multirow{2}{*}{0.5}
& 500  & \textbf{0.0104} ($-0.0020$) & 0.0105 ($-0.0025$) \\
& & 1000 & \textbf{0.0047} ($-0.0008$) & \textbf{0.0047} ($-0.0011$) \\
\midrule
\multirow{4}{*}{II}
& \multirow{2}{*}{0.2}
& 500  & 0.0092 (0.0046) & \textbf{0.0088} ($-0.0019$) \\
& & 1000 & 0.0050 (0.0059) & \textbf{0.0044} ($-0.0006$) \\
\cmidrule(lr){2-5}
& \multirow{2}{*}{0.5}
& 500  & 0.0157 (0.0044) & \textbf{0.0151} ($-0.0018$) \\
& & 1000 & 0.0075 (0.0067) & \textbf{0.0067} (0.0011) \\
\midrule
\multirow{4}{*}{III}
& \multirow{2}{*}{0.2}
& 500  & 0.2951 (0.2006) & \textbf{0.0339} (0.0074) \\
& & 1000 & 0.2472 (0.1846) & \textbf{0.0170} ($-0.0011$) \\
\cmidrule(lr){2-5}
& \multirow{2}{*}{0.5}
& 500  & 0.2864 (0.1726) & \textbf{0.0574} ($-0.0070$) \\
& & 1000 & 0.2673 (0.1900) & \textbf{0.0245} ($-0.0017$) \\
\midrule
\multirow{4}{*}{IV}
& \multirow{2}{*}{0.2}
& 500  & 0.4196 (0.2212) & \textbf{0.0554} ($-0.0088$) \\
& & 1000 & 0.3832 (0.2244) & \textbf{0.0300} (0.0033) \\
\cmidrule(lr){2-5}
& \multirow{2}{*}{0.5}
& 500  & 0.4841 (0.2213) & \textbf{0.1119} ($-0.0044$) \\
& & 1000 & 0.4339 (0.2268) & \textbf{0.0514} (0.0052) \\
\bottomrule
\end{tabular}
\end{table}

In Case I, all methods show similar estimation performance when $K_{\mathrm{true}}=2$. In particular, ZINBCWM shows almost no efficiency loss compared with ZIPCWM even though it additionally estimates the dispersion parameter. In Cases II--IV, the biases of ZIP and ZIPCWM are positive and increase as the degree of overdispersion increases. In Cases III and IV, the biases remain substantial even when the sample size increases. On the other hand, ZINB and ZINBCWM have biases close to zero, and the MSE of ZINBCWM decreases as the sample size increases. ZINBCWM also provides the smallest MSE in all cases involving overdispersion.

When $K_{\mathrm{true}}=3$, the difference between ZIPCWM and
ZINBCWM becomes more apparent. In Cases I and II, the two methods show comparable performance. However, in Cases III and IV, ZIPCWM produces substantially larger positive biases and MSEs. In contrast, ZINBCWM maintains biases close to zero, and its MSE decreases as the sample size increases. These results demonstrate that ZINBCWM provides reliable estimates of the component-specific regression coefficients when overdispersion is present within the count components.

\subsection{Simulation 2: Clustering performance}
\label{sec:sim-clustering}

In the second simulation study, we compare the clustering
performance of ZIPCWM and ZINBCWM under the same four cases.
For each simulated sample, both models are fitted with
$K=2,\ldots,5$, and the number of components is selected using the Bayesian information criterion (BIC; \citealp{schwarz1978})
and the integrated completed likelihood criterion
(ICL; \citealp{biernacki2000}). For a model with $K$ components, BIC is defined as
\[
 \mathrm{BIC}(K)
 =
 -2\ell(\widehat\Theta_K)+p_K\log n,
\]
where $\widehat\Theta_K$ is the parameter estimate under $K$
components and $p_K$ is the number of free parameters. ICL is
defined as
\[
 \mathrm{ICL}(K)
 =
 \mathrm{BIC}(K)
 -2\sum_{i=1}^{n}\sum_{k=1}^{K}
 \widehat z_{ik}\log\widehat z_{ik},
\]
where $\widehat z_{ik}$ is the estimated posterior probability that observation $i$ belongs to component $k$. Thus, ICL additionally incorporates the classification entropy and tends to favor models with well-separated clusters. For each criterion, the value of $K$ that minimizes the corresponding criterion is selected.

For each selected model, observations are assigned to components using the maximum a posteriori rule,
\[
 \widehat Z_i
 =
 \argmax_{1\le k\le K}\widehat z_{ik}.
\]
Clustering performance is evaluated using the adjusted Rand index (ARI; \citealp{hubert1985}) and adjusted mutual information (AMI; \citealp{vinh2010}). Both measures compare the estimated partition with the true component labels while adjusting for agreement by chance, and larger values indicate better clustering performance. Tables~\ref{tab:sim-ari} and~\ref{tab:sim-ami} report the mean ARI and AMI, respectively, over $B=200$ replicated samples. The best value under each simulation setting is shown in bold.

\begin{table}[tbp]
\centering
\caption{Mean ARI for the BIC- and ICL-selected models over
$B=200$ replications.}
\label{tab:sim-ari}
\footnotesize
{\setlength{\tabcolsep}{6pt}{%
\begin{tabular}{lllcccc@{\hspace{10pt}}cccc}
\toprule
Case & $\pi_1$ & $n$
& \multicolumn{4}{c}{$K_{\mathrm{true}}=2$}
& \multicolumn{4}{c}{$K_{\mathrm{true}}=3$} \\
\cmidrule(lr){4-7}\cmidrule(lr){8-11}
& & &
\multicolumn{2}{c}{ZIPCWM}
& \multicolumn{2}{c}{ZINBCWM}
& \multicolumn{2}{c}{ZIPCWM}
& \multicolumn{2}{c}{ZINBCWM} \\
\cmidrule(lr){4-5}\cmidrule(lr){6-7}
\cmidrule(lr){8-9}\cmidrule(lr){10-11}
& & &
BIC & ICL & BIC & ICL &
BIC & ICL & BIC & ICL \\
\midrule
\multirow{4}{*}{I}
& \multirow{2}{*}{0.2}
& 500
& 0.8507 & 0.8507
& \textbf{0.8509} & \textbf{0.8509}
& \textbf{0.9182} & \textbf{0.9182}
& 0.9180 & 0.9180 \\
& & 1000
& \textbf{0.8523} & \textbf{0.8523}
& 0.8522 & 0.8522
& \textbf{0.9217} & \textbf{0.9217}
& 0.9214 & 0.9214 \\
\cmidrule(lr){2-11}
& \multirow{2}{*}{0.5}
& 500
& \textbf{0.8839} & \textbf{0.8839}
& 0.8833 & 0.8833
& \textbf{0.9336} & \textbf{0.9336}
& \textbf{0.9336} & \textbf{0.9336} \\
& & 1000
& 0.8823 & 0.8823
& \textbf{0.8828} & \textbf{0.8828}
& \textbf{0.9351} & \textbf{0.9351}
& 0.9350 & 0.9350 \\
\midrule
\multirow{4}{*}{II}
& \multirow{2}{*}{0.2}
& 500
& 0.8245 & 0.8245
& \textbf{0.8341} & \textbf{0.8341}
& 0.8547 & 0.9018
& \textbf{0.9061} & \textbf{0.9061} \\
& & 1000
& 0.8268 & 0.8291
& \textbf{0.8411} & \textbf{0.8411}
& 0.7585 & 0.9027
& \textbf{0.9092} & \textbf{0.9092} \\
\cmidrule(lr){2-11}
& \multirow{2}{*}{0.5}
& 500
& 0.8582 & 0.8582
& \textbf{0.8614} & \textbf{0.8614}
& 0.9136 & 0.9186
& \textbf{0.9203} & \textbf{0.9203} \\
& & 1000
& 0.8598 & 0.8598
& \textbf{0.8632} & \textbf{0.8632}
& 0.8933 & 0.9197
& \textbf{0.9217} & \textbf{0.9217} \\
\midrule
\multirow{4}{*}{III}
& \multirow{2}{*}{0.2}
& 500
& 0.2468 & 0.2792
& \textbf{0.7704} & \textbf{0.7704}
& 0.3686 & 0.3682
& \textbf{0.8434} & \textbf{0.8434} \\
& & 1000
& 0.2360 & 0.2524
& \textbf{0.7804} & \textbf{0.7804}
& 0.3573 & 0.3589
& \textbf{0.8520} & \textbf{0.8520} \\
\cmidrule(lr){2-11}
& \multirow{2}{*}{0.5}
& 500
& 0.5596 & 0.5566
& \textbf{0.7822} & \textbf{0.7822}
& 0.6608 & 0.6404
& \textbf{0.8339} & \textbf{0.8339} \\
& & 1000
& 0.5609 & 0.5630
& \textbf{0.7846} & \textbf{0.7846}
& 0.6555 & 0.6451
& \textbf{0.8379} & \textbf{0.8379} \\
\midrule
\multirow{4}{*}{IV}
& \multirow{2}{*}{0.2}
& 500
& 0.1290 & 0.0654
& \textbf{0.7441} & \textbf{0.7441}
& 0.2023 & 0.1887
& \textbf{0.8251} & \textbf{0.8251} \\
& & 1000
& 0.1718 & 0.0663
& \textbf{0.7492} & \textbf{0.7492}
& 0.1911 & 0.1797
& \textbf{0.8285} & \textbf{0.8285} \\
\cmidrule(lr){2-11}
& \multirow{2}{*}{0.5}
& 500
& 0.4731 & 0.4290
& \textbf{0.7431} & \textbf{0.7431}
& 0.5438 & 0.5192
& \textbf{0.7863} & \textbf{0.7863} \\
& & 1000
& 0.4943 & 0.4470
& \textbf{0.7516} & \textbf{0.7516}
& 0.5347 & 0.5227
& \textbf{0.7923} & \textbf{0.7923} \\
\bottomrule
\end{tabular}
}
}
\end{table}

\begin{table}[tbp]
\centering
\caption{Mean AMI for the BIC- and ICL-selected models over
$B=200$ replications.}
\label{tab:sim-ami}
\footnotesize
{\setlength{\tabcolsep}{6pt}{%
\begin{tabular}{lllcccc@{\hspace{10pt}}cccc}
\toprule
Case & $\pi_1$ & $n$
& \multicolumn{4}{c}{$K_{\mathrm{true}}=2$}
& \multicolumn{4}{c}{$K_{\mathrm{true}}=3$} \\
\cmidrule(lr){4-7}\cmidrule(lr){8-11}
& & &
\multicolumn{2}{c}{ZIPCWM}
& \multicolumn{2}{c}{ZINBCWM}
& \multicolumn{2}{c}{ZIPCWM}
& \multicolumn{2}{c}{ZINBCWM} \\
\cmidrule(lr){4-5}\cmidrule(lr){6-7}
\cmidrule(lr){8-9}\cmidrule(lr){10-11}
& & &
BIC & ICL & BIC & ICL &
BIC & ICL & BIC & ICL \\
\midrule
\multirow{4}{*}{I}
& \multirow{2}{*}{0.2}
& 500
& \textbf{0.7302} & \textbf{0.7302}
& 0.7299 & 0.7299
& \textbf{0.8731} & \textbf{0.8731}
& 0.8727 & 0.8727 \\
& & 1000
& \textbf{0.7304} & \textbf{0.7304}
& 0.7299 & 0.7299
& \textbf{0.8761} & \textbf{0.8761}
& 0.8756 & 0.8756 \\
\cmidrule(lr){2-11}
& \multirow{2}{*}{0.5}
& 500
& \textbf{0.8161} & \textbf{0.8161}
& 0.8149 & 0.8149
& \textbf{0.8996} & \textbf{0.8996}
& 0.8994 & 0.8994 \\
& & 1000
& 0.8129 & 0.8129
& \textbf{0.8131} & \textbf{0.8131}
& \textbf{0.8999} & \textbf{0.8999}
& 0.8996 & 0.8996 \\
\midrule
\multirow{4}{*}{II}
& \multirow{2}{*}{0.2}
& 500
& \textbf{0.7080} & \textbf{0.7080}
& 0.7075 & 0.7075
& 0.8298 & 0.8565
& \textbf{0.8567} & \textbf{0.8567} \\
& & 1000
& 0.7110 & 0.7122
& \textbf{0.7136} & \textbf{0.7136}
& 0.7730 & 0.8544
& \textbf{0.8577} & \textbf{0.8577} \\
\cmidrule(lr){2-11}
& \multirow{2}{*}{0.5}
& 500
& \textbf{0.7911} & \textbf{0.7911}
& 0.7870 & 0.7870
& 0.8780 & 0.8830
& \textbf{0.8839} & \textbf{0.8839} \\
& & 1000
& \textbf{0.7915} & \textbf{0.7915}
& 0.7870 & 0.7870
& 0.8537 & \textbf{0.8823}
& 0.8818 & 0.8818 \\
\midrule
\multirow{4}{*}{III}
& \multirow{2}{*}{0.2}
& 500
& 0.3570 & 0.3609
& \textbf{0.6218} & \textbf{0.6218}
& 0.4902 & 0.4751
& \textbf{0.7727} & \textbf{0.7727} \\
& & 1000
& 0.3579 & 0.3548
& \textbf{0.6298} & \textbf{0.6298}
& 0.4831 & 0.4781
& \textbf{0.7762} & \textbf{0.7762} \\
\cmidrule(lr){2-11}
& \multirow{2}{*}{0.5}
& 500
& 0.5450 & 0.5524
& \textbf{0.6894} & \textbf{0.6894}
& 0.6007 & 0.5822
& \textbf{0.7778} & \textbf{0.7778} \\
& & 1000
& 0.5431 & 0.5535
& \textbf{0.6893} & \textbf{0.6893}
& 0.5917 & 0.5828
& \textbf{0.7802} & \textbf{0.7802} \\
\midrule
\multirow{4}{*}{IV}
& \multirow{2}{*}{0.2}
& 500
& 0.2795 & 0.2398
& \textbf{0.5870} & \textbf{0.5870}
& 0.3515 & 0.3286
& \textbf{0.7499} & \textbf{0.7499} \\
& & 1000
& 0.3058 & 0.2425
& \textbf{0.5893} & \textbf{0.5893}
& 0.3398 & 0.3218
& \textbf{0.7470} & \textbf{0.7470} \\
\cmidrule(lr){2-11}
& \multirow{2}{*}{0.5}
& 500
& 0.4750 & 0.4564
& \textbf{0.6460} & \textbf{0.6460}
& 0.4875 & 0.4649
& \textbf{0.7292} & \textbf{0.7292} \\
& & 1000
& 0.4787 & 0.4664
& \textbf{0.6517} & \textbf{0.6517}
& 0.4737 & 0.4614
& \textbf{0.7321} & \textbf{0.7321} \\
\bottomrule
\end{tabular}
}
}
\end{table}

In Case I, ZIPCWM and ZINBCWM show almost identical clustering performance under both BIC and ICL. The results are also similar in Case II when $K_{\mathrm{true}}=2$. When
$K_{\mathrm{true}}=3$, however, the clustering performance of
ZIPCWM based on BIC deteriorates as the sample size increases, whereas its ICL-based results remain comparable to those of ZINBCWM. For example, when $\pi_1=0.2$ and $n=1000$, the ARI of ZIPCWM is 0.7585 under BIC and 0.9027 under ICL, compared with 0.909 for ZINBCWM under both criteria.

In Cases III and IV, ZINBCWM substantially outperforms ZIPCWM
in terms of both ARI and AMI regardless of the model-selection criterion. For example, in Case IV with
$K_{\mathrm{true}}=3$, $\pi_1=0.2$, and $n=1000$, the ARI
values of ZIPCWM are 0.1911 and 0.1797 under BIC and ICL,
respectively, whereas ZINBCWM gives 0.8285 under both criteria. The corresponding AMI values are 0.3398 and 0.3218 for ZIPCWM and 0.7470 for ZINBCWM. These results demonstrate that explicitly accounting for within-component overdispersion is important for recovering the underlying cluster structure.

\section{Real data analysis}
\label{sec:real}

In this section, we consider two real-world count datasets to
illustrate the clustering structure identified by ZINBCWM and
its practical interpretation. For each dataset, ZIPCWM and
ZINBCWM are fitted with $K=2,\ldots,5$, and the number of
components is selected using BIC and ICL.
For interpretation of the count regression, we report
$\exp(\widehat\beta_{kj})$ together with each regression
coefficient. This quantity represents the incidence rate ratio (IRR) associated with a one-unit increase in the corresponding covariate, with the other covariates fixed. For standardized continuous covariates, it corresponds to a one-standard-deviation increase.

Since the true cluster memberships are unavailable for real
datasets, clustering performance cannot be directly assessed as in the simulation study. As a supplementary analysis, we compare ZIP, ZINB, ZIPCWM, and ZINBCWM using 10-fold cross-validation. Prediction accuracy is evaluated using the mean absolute error (MAE),
\[
\mathrm{MAE}
=
\frac{1}{n_{\mathrm{test}}}
\sum_{i\in\mathcal I}
|y_i-\widehat y_i|,
\]
where $\mathcal I$ denotes the test fold and $n_{\mathrm{test}}$ is its size. MAE represents the average absolute prediction error, and a smaller value indicates better prediction performance.

\subsection{DoctorVisits data}
\label{subsec:doctor}

For the first real data analysis, we consider the
\texttt{DoctorVisits} data \citep{camerontrivedi1986}, which are available in the \texttt{AER} package in R. The data contain 5,190 individuals, and the response is the number of doctor visits during the previous two weeks. We use age and annual income as continuous covariates; the number of illnesses, reduced-activity days, and the general health questionnaire (GHQ) score as count covariates; and gender, private insurance, government insurance, and chronic conditions with and without activity limitation as categorical covariates. Government insurance indicates whether an individual belongs to either of the two government-supported insurance categories in the original data. Age and income are standardized before model fitting.

The proportion of zero doctor visits is 79.79\%.
Table~\ref{tab:doctor-selection} presents the BIC and ICL values for ZIPCWM and ZINBCWM. Both criteria select $K=5$ within the candidate range for both models. Moreover, ZINBCWM provides smaller BIC and ICL than ZIPCWM at $K=5$.

\begin{table}[ht]
\centering
\caption{BIC and ICL for the DoctorVisits data.
Boldfaced numbers indicate the smallest value in each criterion.}
\label{tab:doctor-selection}
\footnotesize
\setlength{\tabcolsep}{8pt}
\begin{tabular}{ccccc}
\toprule
& \multicolumn{2}{c}{ZIPCWM}
& \multicolumn{2}{c}{ZINBCWM} \\
\cmidrule(lr){2-3}\cmidrule(lr){4-5}
$K$ & BIC & ICL & BIC & ICL \\
\midrule
2 & 115702.00 & 116509.38 & 115651.96 & 116479.63 \\
3 & 105095.53 & 105999.68 & 104957.09 & 105865.42 \\
4 & 101071.53 & 102569.87 & 100936.40 & 102451.57 \\
5 & \textbf{96653.46} & \textbf{98511.31}
  & \textbf{96507.33} & \textbf{98366.51} \\
\bottomrule
\end{tabular}
\end{table}

Table~\ref{tab:doctor-profile} summarizes the five clusters
obtained from ZINBCWM. Cluster 1 is the structural-zero cluster with $\widehat\pi_1=0.2013$. This cluster is characterized by relatively high income, small numbers of illnesses and reduced-activity days, a low GHQ score, a high proportion of private insurance, and a low proportion of government insurance. Thus, this cluster can be interpreted as a relatively healthy, higher-income middle-aged group.

Cluster 2, which has the largest mixing proportion
($\widehat\pi_2=0.3639$), primarily consists of younger
individuals with relatively favorable health profiles. Cluster 3 has the largest number of reduced-activity days and relatively poor health measures, representing a high-health-risk group. Cluster 4 is characterized by the highest age, lowest income, and highest proportion of government insurance, whereas Cluster 5 has relatively large numbers of illnesses and the highest GHQ score, together with high proportions of chronic conditions.

\begin{table}[ht]
\centering
\caption{Estimated cluster profiles of ZINBCWM with $K=5$ for
the DoctorVisits data.}
\label{tab:doctor-profile}
\footnotesize
\setlength{\tabcolsep}{3pt}
\begin{tabular}{lccccc}
\toprule
& Cluster 1 & Cluster 2 & Cluster 3 & Cluster 4 & Cluster 5 \\
\midrule
Number of individuals
& 975 & 1992 & 427 & 963 & 833 \\
Mean doctor visits
& 0.00 & 0.14 & 1.39 & 0.33 & 0.45 \\
Variance of doctor visits
& 0.00 & 0.21 & 3.41 & 0.36 & 0.51 \\
Zero proportion (\%)
& 100.0 & 88.8 & 37.9 & 73.1 & 63.9 \\
\midrule
$\widehat\pi_k$
& 0.2013 & 0.3639 & 0.0818 & 0.1843 & 0.1687 \\
$\widehat\alpha_k$
& -- & 1.0176 & 0.8391 & 0.1425 & 0.0710 \\
\midrule
$\widehat\mu_k$: Age
& 0.3148 & $-0.9173$ & 0.2764 & 1.3152 & 0.0325 \\
$\widehat\mu_k$: Income
& 0.6355 & 0.0739 & $-0.1306$ & $-0.8664$ & 0.0921 \\
\midrule
$\widehat\nu_k$: Number of illnesses
& 0.7863 & 0.9757 & 2.5011 & 1.7697 & 2.2998 \\
$\widehat\nu_k$: Reduced-activity days
& 0.0224 & 0.1346 & 9.3988 & 0.0374 & 0.1953 \\
$\widehat\nu_k$: GHQ score
& 0.1507 & 0.4958 & 3.0808 & 0.6753 & 3.7375 \\
\midrule
$\widehat\rho_k$: Female
& 0.4552 & 0.3851 & 0.5988 & 0.7856 & 0.5637 \\
$\widehat\rho_k$: Private insurance
& 0.6916 & 0.4256 & 0.4044 & 0.1610 & 0.5095 \\
$\widehat\rho_k$: Government insurance
& 0.0378 & 0.0815 & 0.3465 & 0.8225 & 0.2121 \\
$\widehat\rho_k$: Non-limiting chronic condition
& 0.4280 & 0.2443 & 0.3956 & 0.6236 & 0.4786 \\
$\widehat\rho_k$: Limiting chronic condition
& 0.0727 & 0.0460 & 0.3335 & 0.1382 & 0.1922 \\
\bottomrule
\end{tabular}
\end{table}

The estimated dispersion parameters also differ considerably
across clusters. Clusters 2 and 3 have dispersion estimates of 1.0176 and 0.8391, respectively, indicating substantial
overdispersion. In contrast, the estimates for Clusters 4 and 5 are much smaller. This suggests that the amount of residual overdispersion can vary substantially across latent subgroups.

Table~\ref{tab:doctor-regression} presents the estimated
regression coefficients for the four count clusters. In
Cluster 2, one additional reduced-activity day is associated with a 1.9261-fold increase in the expected number of doctor visits, while one additional illness corresponds to a 1.4605-fold increase. Thus, doctor visits in the younger group appear to be particularly associated with immediate illness and activity restriction.
For Cluster 4, a one-standard-deviation increase in age is
associated with a 1.7111-fold increase in the expected number of visits. Private insurance and a limiting chronic condition are also associated with increases of 1.7581 and 1.4819 times, respectively. The association with government insurance differs between clusters: its IRR is 0.6719 in Cluster 3 but 1.5098 in Cluster 5. These results show that the relationships between health-related covariates and doctor visits vary substantially across the latent groups.

\begin{table}[ht]
\centering
\caption{Estimated regression coefficients for the count clusters
in the DoctorVisits data. IRRs are given in parentheses.}
\label{tab:doctor-regression}
\footnotesize
\setlength{\tabcolsep}{3pt}
\begin{tabular}{lcccc}
\toprule
& Cluster 2 & Cluster 3 & Cluster 4 & Cluster 5 \\
\midrule
Intercept
& $-2.4606$ (0.0854)
& $-0.1232$ (0.8841)
& $-2.4422$ (0.0870)
& $-1.0110$ (0.3639) \\
Age
& 0.3435 (1.4099)
& 0.0466 (1.0477)
& 0.5371 (1.7111)
& 0.2972 (1.3461) \\
Income
& 0.0410 (1.0419)
& $-0.0826$ (0.9208)
& 0.0851 (1.0888)
& 0.0382 (1.0389) \\
Number of illnesses
& 0.3788 (1.4605)
& 0.0387 (1.0394)
& 0.1119 (1.1184)
& $-0.0410$ (0.9598) \\
Reduced-activity days
& 0.6555 (1.9261)
& 0.0444 (1.0454)
& 0.1535 (1.1659)
& 0.2740 (1.3152) \\
GHQ score
& $-0.2172$ (0.8047)
& 0.0225 (1.0227)
& 0.1100 (1.1163)
& $-0.0543$ (0.9472) \\
Female
& 0.4363 (1.5470)
& 0.0768 (1.0798)
& $-0.0599$ (0.9419)
& 0.0133 (1.0134) \\
Private insurance
& 0.1591 (1.1725)
& $-0.1191$ (0.8877)
& 0.5643 (1.7581)
& 0.3100 (1.3634) \\
Government insurance
& $-0.1951$ (0.8227)
& $-0.3977$ (0.6719)
& 0.0605 (1.0624)
& 0.4120 (1.5098) \\
Non-limiting chronic condition
& $-0.0516$ (0.9497)
& $-0.0411$ (0.9598)
& 0.3101 (1.3635)
& 0.0378 (1.0385) \\
Limiting chronic condition
& 0.2052 (1.2278)
& $-0.0849$ (0.9186)
& 0.3933 (1.4819)
& 0.2921 (1.3392) \\
\bottomrule
\end{tabular}
\end{table}

As a supplementary assessment, the 10-fold cross-validated MAEs of ZIP, ZINB, ZIPCWM, and ZINBCWM are 0.3971, 0.4042, 0.3851, and 0.3850, respectively. The two CWMs therefore provide smaller MAEs than ZIP and ZINB, while ZIPCWM and ZINBCWM show almost identical performance. This result provides additional evidence that the latent structures identified by the CWMs contain useful information for describing doctor visits.

\subsection{View-T road-safety data}
\label{subsec:viewt}

For the second real data analysis, we consider road-safety data collected from View-T, a national transportation big-data analysis service in Korea. The data consist of 1,024 road segments with one response and seven covariates. The response is the serious accident count, defined as the sum of fatalities and serious injuries. We use road length, traffic volume, free-flow speed, congested speed, average speed, and congestion cost as continuous covariates. All continuous covariates are log-transformed and standardized before model fitting. The number of lanes is reclassified into four categories: 1--2, 3--4, 5--6, and at least 7 lanes, and is treated as a categorical covariate.

The proportion of zero serious accident counts is 41.70\%.
Table~\ref{tab:viewt-selection} presents the model-selection
results. Similar to the DoctorVisits analysis, both BIC and ICL select $K=5$ for ZIPCWM and ZINBCWM within the candidate range. At $K=5$, ZINBCWM has smaller BIC and ICL than ZIPCWM.
Table~\ref{tab:viewt-profile} summarizes the estimated cluster profiles from ZINBCWM. Cluster 1 is the structural-zero cluster, with $\widehat\pi_1=0.0449$. It has the lowest traffic volume and speed levels, and approximately 94\% of the roads have four or fewer lanes. This cluster can therefore be interpreted as a low-traffic small-road group.

\begin{table}[ht]
\centering
\caption{BIC and ICL for the View-T data.
Boldfaced numbers indicate the smallest value in each criterion.}
\label{tab:viewt-selection}
\footnotesize
\setlength{\tabcolsep}{8pt}
\begin{tabular}{ccccc}
\toprule
& \multicolumn{2}{c}{ZIPCWM}
& \multicolumn{2}{c}{ZINBCWM} \\
\cmidrule(lr){2-3}\cmidrule(lr){4-5}
$K$ & BIC & ICL & BIC & ICL \\
\midrule
2 & 19815.42 & 19946.89 & 19808.84 & 19943.66 \\
3 & 17460.00 & 17583.66 & 17452.70 & 17574.55 \\
4 & 15670.03 & 15847.06 & 15666.68 & 15831.73 \\
5 & \textbf{15228.30} & \textbf{15525.38}
  & \textbf{15214.79} & \textbf{15465.09} \\
\bottomrule
\end{tabular}
\end{table}

Cluster 2 consists primarily of large high-speed arterial roads, having the highest traffic volume, free-flow speed, and congested speed, together with a relatively large proportion of roads with at least seven lanes. Cluster 3 has relatively moderate road characteristics and low congestion cost, representing a low-congestion ordinary-road group. Cluster 4 has the highest average speed and the lowest congestion cost and can be viewed as an uncongested high-speed-road group. Cluster 5 has the largest mixing proportion ($\widehat\pi_5=0.4404$), relatively low speeds,
and high congestion cost. It also has the largest mean serious accident count and the smallest zero proportion among the count clusters, representing a chronically congested urban-road group.

\begin{table}[ht]
\centering
\caption{Estimated cluster profiles of ZINBCWM with $K=5$ for
the View-T data.}
\label{tab:viewt-profile}
\footnotesize
\setlength{\tabcolsep}{3pt}
\begin{tabular}{lccccc}
\toprule
& Cluster 1 & Cluster 2 & Cluster 3 & Cluster 4 & Cluster 5 \\
\midrule
Number of road segments
& 48 & 150 & 291 & 87 & 448 \\
Mean serious accident count
& 0.00 & 1.55 & 0.69 & 0.79 & 1.73 \\
Variance of serious accident count
& 0.00 & 4.71 & 0.74 & 1.05 & 4.67 \\
Zero proportion (\%)
& 100.0 & 37.3 & 49.5 & 46.0 & 31.0 \\
\midrule
$\widehat\pi_k$
& 0.0449 & 0.1468 & 0.2816 & 0.0863 & 0.4404 \\
$\widehat\alpha_k$
& -- & 0.0814 & $<0.0001$ & $<0.0001$ & 0.1189 \\
\midrule
$\widehat\mu_k$: Road length
& 0.0925 & $-0.0091$ & $-0.2211$ & 0.1825 & 0.0992 \\
$\widehat\mu_k$: Traffic volume
& $-1.1264$ & 0.7283 & $-0.0032$ & 0.1464 & $-0.1547$ \\
$\widehat\mu_k$: Free-flow speed
& $-0.6395$ & 1.8723 & $-0.1667$ & 0.8522 & $-0.6193$ \\
$\widehat\mu_k$: Congested speed
& $-0.7316$ & 1.5476 & $-0.0976$ & 0.8060 & $-0.5368$ \\
$\widehat\mu_k$: Average speed
& $-0.7390$ & 0.1519 & 0.5976 & 1.3126 & $-0.6146$ \\
$\widehat\mu_k$: Congestion cost
& 0.2134 & 0.6945 & $-0.6660$ & $-1.3816$ & 0.4432 \\
\midrule
$\widehat\rho_k$: 1--2 lanes
& 0.2692 & 0.0532 & 0.3010 & 0.2908 & 0.1555 \\
$\widehat\rho_k$: 3--4 lanes
& 0.6720 & 0.2859 & 0.2567 & 0.3337 & 0.3828 \\
$\widehat\rho_k$: 5--6 lanes
& 0.0587 & 0.2684 & 0.2758 & 0.3057 & 0.3224 \\
$\widehat\rho_k$: 7+ lanes
& 0.0000 & 0.3924 & 0.1665 & 0.0698 & 0.1393 \\
\bottomrule
\end{tabular}

\medskip
\footnotesize
\noindent
The estimates $\widehat\alpha_3$ and $\widehat\alpha_4$ attain the lower bound $10^{-6}$ of the optimization.
\end{table}

The dispersion estimates are generally smaller than those
obtained for the DoctorVisits data. Mild overdispersion is
observed in Clusters 2 and 5, with $\widehat\alpha_2=0.0814$ and $\widehat\alpha_5=0.1189$, respectively. In contrast, the estimates for Clusters 3 and 4 are close to zero. Hence, ZINBCWM can simultaneously accommodate clusters with and without substantial overdispersion.

Table~\ref{tab:viewt-regression} presents the estimated
regression coefficients. Road length has a positive coefficient in all count clusters and has the largest association in Cluster 5, where a one-standard-deviation increase in log-transformed road length corresponds to an IRR of 2.7145. The effects of traffic volume and free-flow speed vary across clusters. The IRR of traffic volume is 0.6261 in Cluster 2 but 1.6008 in Cluster 5. Similarly, free-flow speed has positive coefficients in Clusters 2 and 3 but negative coefficients in Clusters 4 and 5, with IRRs of 0.6294 and 0.4481, respectively.
For Cluster 3, congestion cost has an IRR of 1.4403, while roads with at least seven lanes in Cluster 4 have an expected serious accident count approximately 2.6 times that of roads with 1--2 lanes. These results demonstrate that the associations between road characteristics and serious accident counts vary considerably across road types.

\begin{table}[ht]
\centering
\caption{Estimated regression coefficients for the count clusters
in the View-T data. IRRs are given in parentheses.}
\label{tab:viewt-regression}
\footnotesize
\setlength{\tabcolsep}{3pt}
\begin{tabular}{lcccc}
\toprule
& Cluster 2 & Cluster 3 & Cluster 4 & Cluster 5 \\
\midrule
Intercept
& $-6.8877$ (0.0010)
& $-0.1314$ (0.8768)
& 0.1443 (1.1553)
& $-0.3911$ (0.6763) \\
Road length
& 0.2057 (1.2284)
& 0.1012 (1.1065)
& 0.1407 (1.1511)
& 0.9986 (2.7145) \\
Traffic volume
& $-0.4682$ (0.6261)
& 0.0331 (1.0336)
& 0.0219 (1.0222)
& 0.4705 (1.6008) \\
Free-flow speed
& 0.3253 (1.3844)
& 0.0462 (1.0473)
& $-0.4629$ (0.6294)
& $-0.8027$ (0.4481) \\
Congested speed
& 3.6385 (38.0346)
& 0.0395 (1.0403)
& 0.0697 (1.0721)
& $-0.1031$ (0.9021) \\
Average speed
& $-3.1154$ (0.0444)
& 0.1692 (1.1844)
& $-0.1066$ (0.8989)
& $-0.0962$ (0.9083) \\
Congestion cost
& 1.1420 (3.1331)
& 0.3649 (1.4403)
& 0.2015 (1.2232)
& $-0.0280$ (0.9724) \\
3--4 lanes
& 0.5241 (1.6889)
& $-0.0726$ (0.9300)
& 0.1320 (1.1411)
& $-0.1412$ (0.8683) \\
5--6 lanes
& 0.4687 (1.5978)
& $-0.2929$ (0.7461)
& 0.2570 (1.2931)
& 0.0273 (1.0277) \\
7+ lanes
& 1.1589 (3.1864)
& $-0.0885$ (0.9153)
& 0.9688 (2.6348)
& 0.2156 (1.2406) \\
\bottomrule
\end{tabular}
\end{table}

As a supplementary assessment, the 10-fold cross-validated MAEs of ZIP, ZINB, ZIPCWM, and ZINBCWM are 0.9839, 0.9809, 0.9675, and 0.9664, respectively. ZINBCWM achieves the smallest MAE among all methods, indicating the best predictive performance. In addition, both CWMs outperform ZIP and ZINB, suggesting that the latent clustering structure is useful for describing serious accident counts. These results further support the advantage of ZINBCWM in accounting for heterogeneous count data.

\section{Discussion}\label{sec:discussion}

In this paper, we propose a zero-inflated negative binomial
cluster-weighted model for count data with excess zeros,
overdispersion, and mixed-type covariates. The proposed model
represents the structural-zero population by a separate component and allows the remaining count components to have distinct regression relationships, dispersion levels, and covariate distributions. Thus, the model can simultaneously account for structural zeros, within-cluster overdispersion, and heterogeneity in both the response--covariate relationships and covariate profiles.

In contrast to existing generalized and zero-inflated CWMs
\citep{Ingrassia2015,Mazza2018,Pocuca2020,Olobatuyi2022},
the use of component-specific negative binomial distributions
allows residual overdispersion to be accommodated within each
count cluster rather than through additional Poisson components.
Moreover, the identifiability results cover mixed-type covariates and do not require the presence of continuous covariates, which broadens the range of data structures to which the proposed framework can be applied.

Despite these advantages, ZINBCWM still relies on parametric
distributional assumptions. In particular, the negative binomial distribution is assumed for each count component, while Gaussian, Poisson, and multinomial distributions are used for continuous, count, and categorical covariates, respectively. Although the negative binomial distribution provides greater flexibility than the Poisson distribution for overdispersed count data, its dispersion structure is determined by a specific parametric form.
Similarly, the assumed covariate distributions may be restrictive when the within-cluster distributions exhibit more complicated features. In addition, the number of parameters increases with the dimension of the covariates and the number of components, which may limit direct application to high-dimensional data.

A particularly useful extension is to replace the parametric
count distribution with a semiparametric Poisson mixture
distribution. Specifically, the conditional probability mass
function for the $k$th count component can be represented as
\[
f_k(y\mid\xb)
=
\int_0^\infty
f_{\mathrm P}\{y;u\lambda_k(\xb)\}\,dG_k(u),
\]
where $G_k$ is a mixing distribution on the positive real line.
The Poisson distribution is obtained when $G_k$ is degenerate at one, whereas the negative binomial distribution results from a gamma mixing distribution \citep{LandMcCallNagin1996}. Rather than specifying a particular parametric family for $G_k$, it can be left unspecified and estimated using the nonparametric maximum
likelihood estimator (NPMLE). Such an approach follows the general NPMLE framework for mixture distributions
\citep{KieferWolfowitz1956,Lindsay1995} and would allow the count distribution within each cluster to adapt flexibly to the observed data. Related NPMLE-based semiparametric approaches have been successfully applied to robust linear regression and accelerated failure time models to reduce the risk of distributional misspecification \citep{SeoOh2024,OhLeeKangSeo2026}. Developing
this semiparametric extension, together with its identifiability and estimation theory, would therefore provide a natural generalization of both ZIPCWM and ZINBCWM.

\begingroup
\small
\setstretch{1.0}
\setlength{\bibsep}{4pt}
\IfFileExists{apa.bst}{\bibliographystyle{apa}}{\bibliographystyle{apalike}}
\bibliography{ref}
\endgroup

\clearpage

\section*{Appendix A: Proof of Lemma \ref{lem:count}}

For $f\in\mathcal F$, consider its probability generating function
\[
 G_f(t)=\sum_{y=0}^{\infty}t^y f(y),
\]
for positive real $t$ at which the series converges.
This transformation is linear on finite linear combinations
over their common domain of convergence and is one-to-one,
since its power-series coefficients determine the probability
mass function.

For the negative binomial distribution in~\eqref{eq:nb},
\begin{equation}\label{eq:nb-pgf}
 G_{\mathrm{NB}}(t;\lambda,\alpha)
 =\{1+\alpha\lambda(1-t)\}^{-1/\alpha},
 \qquad
 0<t<T(\lambda,\alpha),
\end{equation}
where
\[
 T(\lambda,\alpha)
 =1+\frac{1}{\alpha\lambda}.
\]
For the Poisson distribution and the point mass at zero,
\begin{equation}\label{eq:poisson-zero-pgf}
 G_{\mathrm P}(t;\lambda)
 =\exp\{\lambda(t-1)\},
 \qquad
 G_{\delta_0}(t)=1,
 \qquad t>0.
\end{equation}
Thus the positive-real convergence domains are
$(0,T(\lambda,\alpha))$ for a negative binomial distribution
and $(0,\infty)$ for a Poisson distribution or $\delta_0$.

We define a total order $\prec$ on $\mathcal F$ as follows.
All negative binomial distributions precede all Poisson
distributions, and all Poisson distributions precede $\delta_0$.
Among negative binomial distributions, the distribution with
smaller $T(\lambda,\alpha)$ precedes the other; when the
endpoints coincide, the distribution with smaller $\alpha$
precedes the other. Among Poisson distributions, the distribution
with larger $\lambda$ precedes the other.
The negative binomial ordering distinguishes all parameter pairs,
because $T(\lambda,\alpha)$ and $\alpha$ uniquely determine
$\lambda$.

First consider two negative binomial distributions with parameters
$(\lambda_j,\alpha_j)$ and endpoints
$T_j=T(\lambda_j,\alpha_j)$, $j=1,2$.
If $T_1<T_2$, then
$G_{\mathrm{NB}}(t;\lambda_1,\alpha_1)\to\infty$
as $t\uparrow T_1$, whereas
$G_{\mathrm{NB}}(t;\lambda_2,\alpha_2)$ has a finite limit.
Hence
\[
 \lim_{t\uparrow T_1}
 \frac{G_{\mathrm{NB}}(t;\lambda_2,\alpha_2)}
      {G_{\mathrm{NB}}(t;\lambda_1,\alpha_1)}
 =0.
\]
If $T_1=T_2$ and $\alpha_1<\alpha_2$, then
$\alpha_1\lambda_1=\alpha_2\lambda_2$, and
\[
 \frac{G_{\mathrm{NB}}(t;\lambda_2,\alpha_2)}
      {G_{\mathrm{NB}}(t;\lambda_1,\alpha_1)}
 =
 \{1+\alpha_1\lambda_1(1-t)\}^{
       1/\alpha_1-1/\alpha_2}
 \longrightarrow0
 \qquad\text{as }t\uparrow T_1.
\]
The exponent is strictly positive, so the limit follows.

Next, for any fixed negative binomial parameters
$(\lambda,\alpha)$ and Poisson mean $\lambda_{\mathrm P}>0$,
the Poisson generating function and $G_{\delta_0}$ remain
finite at $T(\lambda,\alpha)$, while the negative binomial
generating function diverges. Therefore,
\[
 \lim_{t\uparrow T(\lambda,\alpha)}
 \frac{G_{\mathrm P}(t;\lambda_{\mathrm P})}
      {G_{\mathrm{NB}}(t;\lambda,\alpha)}
 =
 \lim_{t\uparrow T(\lambda,\alpha)}
 \frac{G_{\delta_0}(t)}
      {G_{\mathrm{NB}}(t;\lambda,\alpha)}
 =0.
\]

Finally, if $\lambda_1>\lambda_2>0$, then
\[
 \frac{G_{\mathrm P}(t;\lambda_2)}
      {G_{\mathrm P}(t;\lambda_1)}
 =
 \exp\{(\lambda_2-\lambda_1)(t-1)\}
 \longrightarrow0
 \qquad\text{as }t\to\infty.
\]
For every $\lambda>0$,
\[
 \frac{G_{\delta_0}(t)}
      {G_{\mathrm P}(t;\lambda)}
 =
 \exp\{-\lambda(t-1)\}
 \longrightarrow0
 \qquad\text{as }t\to\infty.
\]

For every ordered pair $f_1\prec f_2$, the convergence
domain of $G_{f_1}$ is contained in that of $G_{f_2}$.
Moreover, $G_{f_2}(t)/G_{f_1}(t)\to0$ at a boundary point
depending only on $f_1$: its finite upper convergence
endpoint if $f_1$ is negative binomial, and $+\infty$
if $f_1$ is Poisson. These properties verify the
transform-ratio criterion of \citet{Teicher1963}.
It follows that finite mixtures of distinct members of
$\mathcal F$ with strictly positive mixing weights are
identifiable up to a permutation of their components.

\section*{Appendix B: Proof of Theorem \ref{thm:ident}}

Write
\[
 h_k(\xb)=\pi_k g_k(\xb;\etab_k),
 \qquad
 h'_j(\xb)=\pi'_j g_j(\xb;\etab'_j).
\]
Primes denote quantities under $\Theta'$. By (A1)--(A2),
all $h_k$ and $h'_j$ are strictly positive on the common
covariate support. Equality of the joint distributions implies
$p_X=p'_X$ and hence equality of the corresponding conditional
mixtures in~\eqref{eq:conditional}. When continuous covariates
are present, for each fixed $(y,\db,\wb)$ the joint density is
continuous in $\qb$. Since a continuous function that vanishes
Lebesgue-almost everywhere vanishes everywhere, and the remaining
coordinates are countable, the assumed equality may be used for
all $y$ and all $\xb\in\calX$. When all covariates are discrete,
this follows directly from the counting measure.

Fix a categorical configuration $\wb$; when $p_W=0$, there is
only one such configuration. If two count components have
different dispersions, their negative binomial distributions are
distinct at every covariate value. For a pair $k<\ell$ with
$\alpha_k=\alpha_\ell$, equality of the two distributions requires
equality of their means, or equivalently
\[
 L_{k\ell,\wb}(\ub)
 :=
 \xi_k(\ub,\wb)-\xi_\ell(\ub,\wb)
 =
 a_k(\wb)-a_\ell(\wb)
 +\ub^\top(\bb_k-\bb_\ell)
 =0.
\]
For every equal-dispersion pair, $L_{k\ell,\wb}$ is a nonzero
polynomial. When $p_W=0$, this follows from pairwise distinctness
of $(\betab_k,\alpha_k)$; when $p_W>0$, equal dispersions imply
$\bb_k\ne\bb_\ell$. The same argument applies to the primed
representation.

Let $H_{\wb}(\ub)$ be the product of these collision polynomials
over all equal-dispersion pairs in both representations, with an
empty product defined as one. Since each factor is a nonzero
polynomial and the polynomial ring over $\R$ has no zero divisors,
$H_{\wb}$ is nonzero. Hence, by Lemma~\ref{lem:polynomial},
\[
 \calS^\circ_{\wb}
 =
 \{\ub\in\calS:H_{\wb}(\ub)\ne0\}
\]
is nonempty. At every $\ub\in\calS^\circ_{\wb}$, the count
distributions within each representation are pairwise distinct,
and $\delta_0$ is distinct from all of them. Since all conditional
mixing weights are positive, Lemma~\ref{lem:count} applied to the
conditional mixture gives $K'=K$ and a pointwise matching of the
response components and their conditional weights, with the
structural-zero component matched to itself.

Fix $k\in\{2,\ldots,K\}$ and define
\[
 J_k
 =
 \{j\in\{2,\ldots,K\}:\alpha'_j=\alpha_k\}.
\]
For every $\ub\in\calS^\circ_{\wb}$, pointwise identifiability
implies that at least one $j\in J_k$ satisfies
\[
 \xi_k(\ub,\wb)=\xi'_j(\ub,\wb).
\]
Thus
\[
 P_{k,\wb}(\ub)
 =
 \prod_{j\in J_k}
 \{\xi_k(\ub,\wb)-\xi'_j(\ub,\wb)\}
\]
satisfies $P_{k,\wb}(\ub)=0$ whenever
$H_{\wb}(\ub)\ne0$. The second assertion of
Lemma~\ref{lem:polynomial} therefore gives
$P_{k,\wb}(\ub)\equiv0$. Since the polynomial ring has no zero
divisors, one factor must be identically zero. Hence, for some
$j\in J_k$,
\begin{equation}\label{eq:whole-function-match}
 a_k(\wb)=a'_j(\wb),
 \qquad
 \bb_k=\bb'_j,
 \qquad
 \alpha_k=\alpha'_j.
\end{equation}

When $p_W=0$, these equalities identify the full regression vector
and dispersion. When $p_W>0$, the pairs $(\bb_k,\alpha_k)$ are
pairwise distinct and do not depend on $\wb$, so the matching
index $j$ is unique and must be the same for every categorical
configuration. The correspondence is also injective, because two
distinct components matched to the same primed component would
have identical $(\bb_k,\alpha_k)$ pairs. Since both
representations contain $K-1$ count components, the correspondence
is bijective and defines a permutation $\tau$ of the count
labels. Set $\tau(1)=1$. Then
\[
 \alpha_k=\alpha'_{\tau(k)},
 \qquad
 \bb_k=\bb'_{\tau(k)},
 \qquad
 a_k(\wb)=a'_{\tau(k)}(\wb)
 \quad\text{for all }\wb\in\calW.
\]
If $p_W>0$, evaluating the last equality at the all-reference
configuration gives
$\beta_{k0}=\beta'_{\tau(k)0}$. Changing one categorical variable
at a time to each nonreference level identifies the corresponding
entries of $\betab_{k,W}$. Together with
$\bb_k=\bb'_{\tau(k)}$, this yields
\[
 \betab_k=\betab'_{\tau(k)},
 \qquad k=2,\ldots,K.
\]

For $\ub\in\calS^\circ_{\wb}$, the fixed matching and
Lemma~\ref{lem:count} also give
\[
 \omega_k(\ub,\wb)
 =
 \omega'_{\tau(k)}(\ub,\wb),
 \qquad k=1,\ldots,K.
\]
Since $p_X=p'_X$,
\begin{equation}\label{eq:h-on-regular}
 h_k(\ub,\wb)
 =
 h'_{\tau(k)}(\ub,\wb),
 \qquad
 \ub\in\calS^\circ_{\wb}.
\end{equation}
The complement of $\calS^\circ_{\wb}$ cannot in general be
discarded because collision points involving discrete covariates
may have positive probability. For fixed $\wb$, consider
\begin{align}\label{eq:log-ratio}
 R_{k,\wb}(\qb,\db)
 &:=
 \log
 \frac{h_k(\qb,\db,\wb)}
      {h'_{\tau(k)}(\qb,\db,\wb)}
 \\
 &=
 C_{k,\wb}
 +\qb^\top\Ab_k\qb
 +\cb_k^\top\qb
 +\sum_{j=1}^{p_D}
 d_j\log
 \frac{\nu_{kj}}
      {\nu'_{\tau(k)j}},
 \nonumber
\end{align}
where $C_{k,\wb}$ is constant in $(\qb,\db)$. The Gaussian
log-density ratio is quadratic in $\qb$, the Poisson factorial
terms cancel, and the categorical probabilities contribute only
to the constant term. Thus $R_{k,\wb}$ is a polynomial in
$\ub=(\qb^\top,\db^\top)^\top$.

By~\eqref{eq:h-on-regular},
$R_{k,\wb}(\ub)=0$ whenever $H_{\wb}(\ub)\ne0$. Applying the
second assertion of Lemma~\ref{lem:polynomial} again gives
$R_{k,\wb}(\ub)\equiv0$, and hence
\[
 h_k(\ub,\wb)=h'_{\tau(k)}(\ub,\wb)
 \qquad\text{for all }\ub\in\calS.
\]
Since $\wb$ was arbitrary,
\[
 h_k(\xb)=h'_{\tau(k)}(\xb),
 \qquad
 \xb\in\calX,
 \qquad k=1,\ldots,K.
\]
The same conclusion holds in the zero-dimensional case, where
the log ratio is constant.

Finally, let $\mu_X$ denote the product reference measure on
$\calX$. Integrating the preceding equality and using the
normalization of each component covariate distribution gives
\[
 \pi_k
 =
 \int_{\calX}h_k(\xb)\,d\mu_X(\xb)
 =
 \int_{\calX}h'_{\tau(k)}(\xb)\,d\mu_X(\xb)
 =
 \pi'_{\tau(k)}.
\]
Since the mixing proportions are positive,
\[
 g_k(\xb;\etab_k)
 =
 g_{\tau(k)}(\xb;\etab'_{\tau(k)}),
 \qquad
 \xb\in\calX.
\]
Marginalization then identifies the Gaussian means and covariance
matrices, the Poisson means, and the categorical probabilities, so
that
\[
 \etab_k=\etab'_{\tau(k)},
 \qquad k=1,\ldots,K.
\]
\end{document}